\documentclass[aps,prl,nobibnotes,superscriptaddress,showpacs,floatfix,reprint,longbibliography,footinbib]{revtex4-2}

\usepackage{natbib}
\usepackage[english]{babel}
\usepackage{letltxmacro}
\usepackage{latexsym}
\LetLtxMacro{\ORIGselectlanguage}{\selectlanguage}
\makeatletter
\DeclareRobustCommand{\selectlanguage}[1]{%
  \@ifundefined{alias@\string#1}
    {\ORIGselectlanguage{#1}}
    {\begingroup\edef\x{\endgroup
       \noexpand\ORIGselectlanguage{\@nameuse{alias@#1}}}\x}%
}
\newcommand{\definelanguagealias}[2]{%
  \@namedef{alias@#1}{#2}%
}
\makeatother
\definelanguagealias{en}{english}
\definelanguagealias{English}{english}
\usepackage{graphicx}
\usepackage{amsmath}
\usepackage{amsfonts}
\usepackage{amssymb}
\usepackage{bm}
\usepackage{color}
\usepackage[percent]{overpic}
\usepackage{soul} 
\usepackage{amssymb}
\usepackage{wasysym}
\usepackage{dsfont}
\usepackage{float}
\usepackage[mathscr]{euscript}
\usepackage{lipsum}
\usepackage{hyperref}
\hypersetup{
    pdfstartview={FitH},    
    colorlinks=true,       
    linkcolor=blue,          
    citecolor=blue,        
    filecolor=magenta,      
    urlcolor=blue           
}

\usepackage{pdfpages}
\usepackage{pgffor}
\makeatletter
\AtBeginDocument{\let\LS@rot\@undefined}
\makeatother

\newcommand{\be}{\begin{equation}}
\newcommand{\ee}{\end{equation}}
\newcommand{\bea}{\begin{eqnarray}}
\newcommand{\eea}{\end{eqnarray}}

\usepackage{graphicx}
\usepackage[colorinlistoftodos]{todonotes}
\usepackage{verbatim}
\usepackage[normalem]{ulem}
\usepackage{enumitem}
\usepackage{multirow}
\usepackage{lipsum}

\begin{document}

\title{Three-body interactions in Rydberg lattices}

\author{Rhine Samajdar}
\affiliation{Department of Electrical and Computer Engineering, Princeton University, Princeton, NJ 08544, USA}
\affiliation{Department of Physics, Princeton University, Princeton, NJ 08544, USA}
\author{Mikhail D. Lukin}
\affiliation{Department of Physics, Harvard University, Cambridge, MA 02138, USA}
\author{Valentin Walther}
\email{vwalther@purdue.edu}
\affiliation{Department of Chemistry, Purdue University, West Lafayette, IN 47907, USA}
\affiliation{Department of Physics and Astronomy, Purdue University, West Lafayette, IN 47907, USA}
\affiliation{Department of Physics, Harvard University, Cambridge, MA 02138, USA}

\begin{abstract}
Programmable arrays of neutral Rydberg atoms are one of the leading platforms today for scalable quantum simulation and computation. In these systems, the dipole-dipole interactions between the individual atoms, or qubits, typically result in binary---i.e., two-body---couplings. In this work, we develop an experimentally accessible scheme for engineering three-body interactions in Rydberg lattices. Such strong three-body couplings can fundamentally modify the underlying physics compared to systems with only two-body interactions: we demonstrate this, in particular, by systematically investigating the effective many-body Hamiltonian and its emergent quantum phases. This capability paves the way for the quantum simulation of a broader class of correlated models of condensed matter and high-energy physics.
\end{abstract}

\maketitle

\textit{Introduction.---}The Coulomb interaction, which underlies most of condensed matter and atomic physics as well as quantum chemistry, is a \textit{binary} force, acting between pairs of particles. Beyond such ubiquitous pairwise forces, hierarchies of \textit{multibody} interactions---even if absent at a fundamental level---can emerge when some of a system's dynamical degrees of freedom are eliminated to obtain effective low-energy descriptions~\cite{Hammer2013}. Prominent  examples in this regard include polarization forces, such as the Casimir-Polder~\cite{power2001} and van der Waals forces~\cite{salam2016}. Initially discovered many decades earlier~\cite{london1937, bade1957, renne1967}, these nonadditive forces continue to be investigated today in dense ensembles of polarizable organic and inorganic molecules, where many-body polarization affects the stability of phases~\cite{distasio2012, shtogun2010, donchev2006, cao1992}.

\begin{figure}[t]
\includegraphics[width=\linewidth]{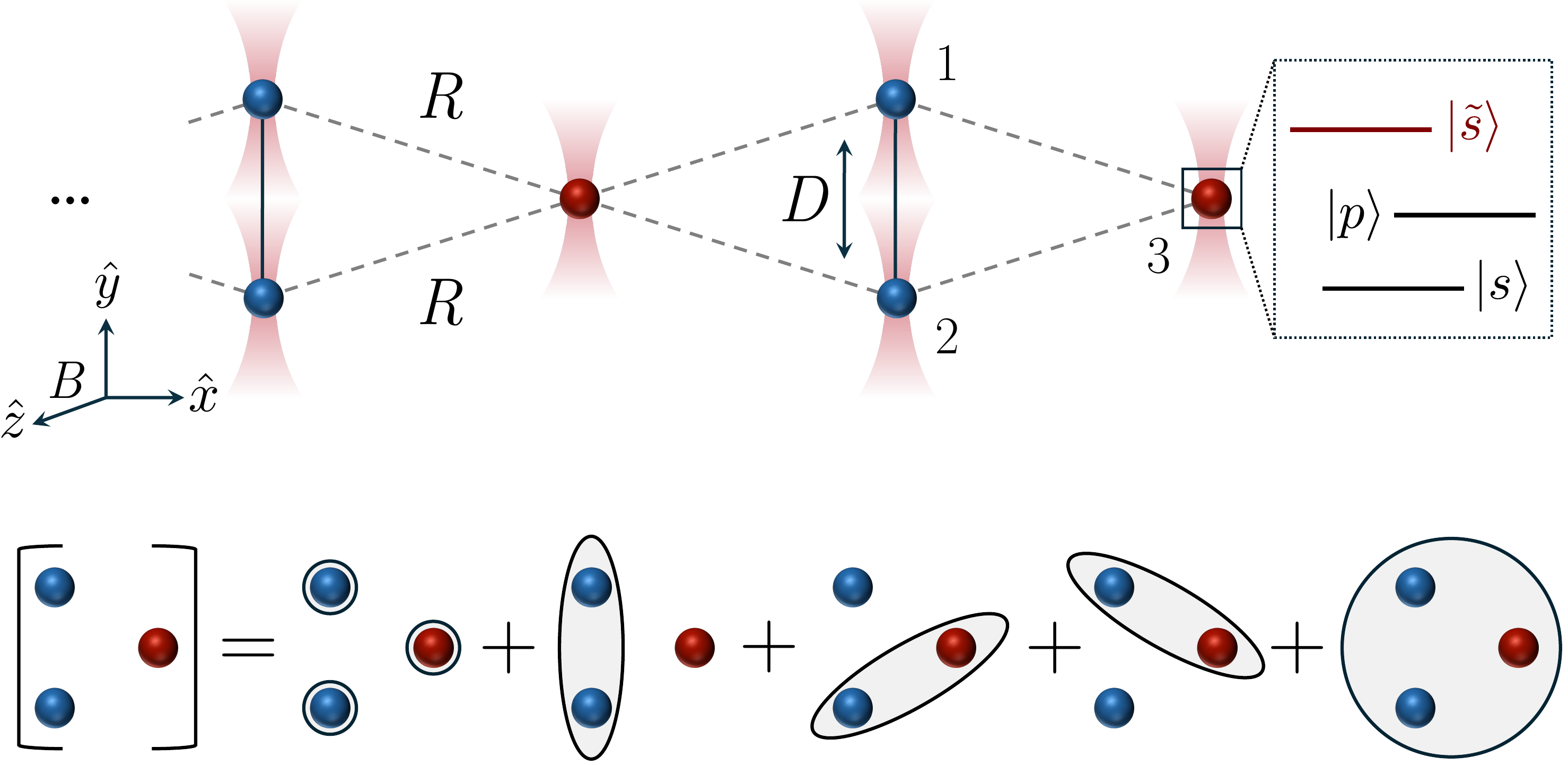}
\caption{Neutral atoms held in optical tweezers are arrayed in a quasi-one-dimensional (1D) geometry and excited to high-lying Rydberg states, $|s\rangle$ and $|\tilde{s}\rangle$. We use a magnetic bias field to split the Zeeman levels and  an AC Stark shift to tune $|\tilde{s}\rangle$ energetically relative to the remaining states. As schematically shown in the lower panel, the net interaction between three atoms consists of single-site components, pairwise terms, as well as intrinsically three-body interactions.}
\label{fig:fig1}
\end{figure}

The presence of multibody (i.e., non-pairwise) interactions originating from van der Waals forces among single atoms has been long appreciated~\cite{axilrod1943} and investigated for ground states of few-atomic systems~\cite{Bell1965, Yan2016, Yan2018, Yan2020}. When atoms are excited to their high-lying Rydberg levels, the availability of a large number of states enhances the importance of these multibody interactions. The first signatures of this physics were observed in dense samples~\cite{Carroll2006, younge2009, faoro2015} and explained using a cluster-expansion theory~\cite{pohl2009}; the driving mechanism thereof was identified as a resonant energy transfer~\cite{faoro2015, tretyakov2017, ryabtsev2018}. Extending this mechanism to include motional effects~\cite{gambetta2020}, the three-body interactions lead to trimer bound states~\cite{kiffner2013}, to more general principal quantum numbers~\cite{Cheinet2020}, as well as to energy transfer involving four atoms~\cite{Gurian2012}. Microwave control of three-body interactions was explored in a simplified model in Ref.~\onlinecite{sevincli2014}.

Recently, programmable arrays of Rydberg atoms have emerged as versatile platforms for quantum simulation~\cite{altman2021quantum}, with the controllable many-body interactions in these systems providing unprecedented access to rich many-body physics~\cite{Browaeys2020}. While these Rydberg simulators natively involve Ising-~\cite{Ebadi.2021,scholl2021,Samajdar_2020} or XY-type~\cite{leseleuc2019,Chen_2023} two-spin interactions, more general  quantum simulation applications naturally call for greater flexibility and versatility. This requires, in particular, the ability to engineer terms corresponding to three-body interactions and beyond. Such three-body terms play important roles in many contexts, including condensed matter physics and materials science; e.g., the scalar spin chirality $\textbf{S}_1\cdot (\textbf{S}_2 \times \textbf{S}_3)$, which arises in a perturbative expansion of the Hubbard model~\cite{sen1995large}, is known to be essential for obtaining certain species of correlated spin liquid states~\cite{nielsen2013local}.

Previous proposals for the (static) realization of three-body interactions have included cold dipolar molecules in optical lattices~\cite{buechler2007b} and in free space~\cite{petrov2014}, nuclear spins~\cite{peng2009}, circuit-QED platforms~\cite{hafezi2014}, and Bose--Hubbard models in the hard-core limit~\cite{mazza2010, daley2014}. More recently, a method for generating multibody interactions via Floquet driving was proposed~\cite{5qhh-322q} and experimentally implemented~\cite{geim2026}. However, Floquet protocols generically induce heating, which limits access to long-time dynamics. In this Letter, we propose a new \textit{time-independent} method for realizing multibody interactions in Rydberg tweezer arrays~\cite{endres2016atom}.
Remarkable experimental progress~\cite{bluvstein2021quantum}  allows for the near-arbitrary spatial placement of qubits, and we investigate the opportunity this offers---together with site-controlled detunings---to engineer dominant three-atom interactions in a lattice system. We also systematically analyze the quantum phases and phase transitions of the resultant (and rather unique) spin model to highlight how three-body interactions can drive the emergence of new phenomena, in departure from the physics with two-body interactions alone.

\textit{Interactions between Rydberg atoms.---}We consider a lattice of $N$ neutral atoms that can be optically excited to Rydberg states. The atoms can be arbitrarily arranged in optical tweezers, but their positions $\vec{R}_i$ are assumed to be close enough for retardation effects to be negligible. The full Hamiltonian of the system is given by ${\mathcal{H}}$\,$=$\,${\mathcal{H}}_0 $\,$+$\,$ \sum_{i \neq j} \mathcal{V}_{ij}$, where ${\mathcal{H}}_0$ describes the unperturbed atomic levels and $\mathcal{V}_{ij}$ is the pairwise dipole-dipole interaction, 
$
\mathcal{V}_{ij} = (1/{4\pi \epsilon_0}) ( {\vec{d}_i \cdot\vec{d}_j}/{|\vec{R}_{ij}|^3} - {(\vec{d}_i\cdot \vec{R}_{ij}) (\vec{d}_j\cdot \vec{R}_{ij})}/{|\vec{R}_{ij}|^5} ),  
$
with $\vec{d}_i$ denoting the dipole operator for atom $i$, and $\vec{R}_{ij}$\,$\equiv$\,$\vec{R}_i$\,$-$\,$\vec{R}_j$. We consider a planar lattice, for which it is convenient to choose the quantization axis out of the plane along the $\hat{z}$ direction (see Fig.~\ref{fig:fig1}). 
Compared to the case of only two interacting atoms (where the quantization axis can be favorably chosen along the interatomic axis to exploit permutation and reflection symmetries~\cite{weber2017}), here, we need to explicitly account for the relative orientations of all atoms.
Consequently, only the parity of the total angular momentum projection $M\equiv\sum_i m_i$ remains a constant of motion. 

\begin{figure}[t]
\begin{center}
 \includegraphics[width=0.9\linewidth]{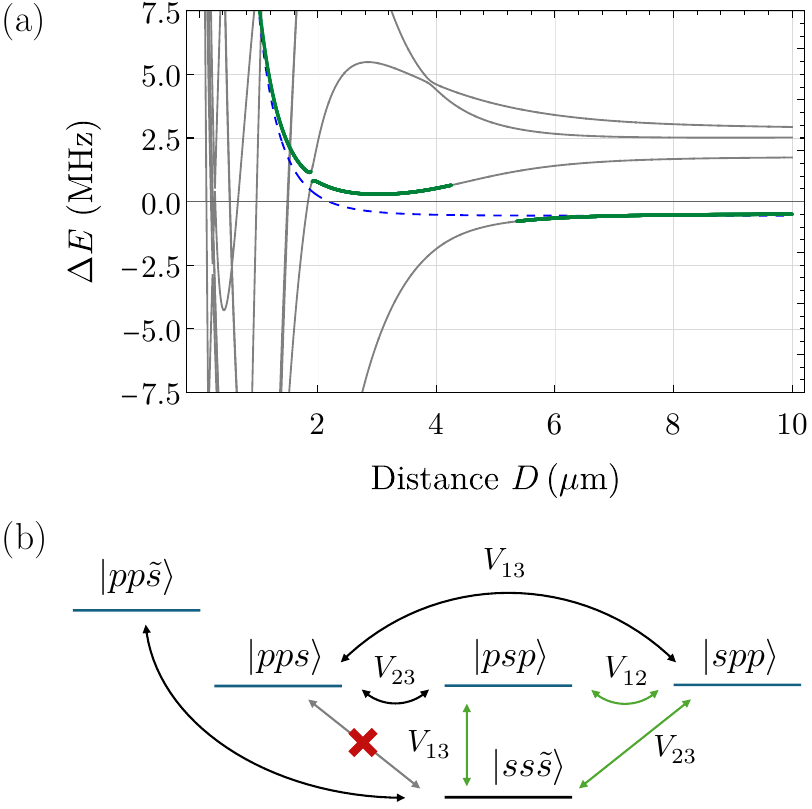}
\end{center}
\caption{(a) The potential energy surfaces of three interacting Rydberg atoms include binary contributions (blue, dashed line). In the vicinity of an avoided crossing of two coupling potentials, three-body interactions can be dominant. The points highlighted in green mark a large overlap ($> 90\%$) with the states $|ss\tilde{s}\rangle$. The potential surfaces are numerically calculated for varying distances $D$ between atoms 1 and 2 while the remaining distances are held constant at 10 $\mu$m. (b) The interaction can be understood in a simplified model involving the states $|s\rangle$, $|\tilde{s}\rangle$, and $|p\rangle$ (taken to be $\vert 37S_{-1/2}\rangle$, $\vert 38S_{-1/2}\rangle$, and  $\vert 37P_{-3/2}\rangle$, respectively, in this example). A near-F\"orster resonance between $|s\tilde{s}\rangle$ and $|pp\rangle$ enhances the three-body interaction, while the state $|pp\tilde{s}\rangle$ is far off resonant.}
\label{fig:fig2}
\end{figure}

In experiments, the atoms are typically all excited to large-principal-quantum-number $s$ states, called ``target states'', whose interactions in turn drive the many-body physics. In most cases, the conditions are chosen such that these interactions strengths are small compared to the single-atom energy-level spacings. From the symmetry of the dipole-dipole interactions, it is immediately clear that the first-order perturbative correction to the two-atom state $\rvert ss \rangle$ is zero: this contribution vanishes because dipolar transitions change the parity of a given state. The second-order perturbation of the many-body potential energy surface then arises from virtual pairwise transitions to nearby states (e.g., $\rvert ss\rangle$\,$\rightarrow \rvert p p' \rangle$\,$\rightarrow \rvert s s \rangle$) separated by a F\"orster defect $\delta$~\cite{browaeys2016experimental}. These processes lead to the well-known and much-used effective interaction of the van der Waals type $V^{(2)}_{ij} = \sum_{i\leq j} C^{}_6/R_{ij}^6$. 

Carrying out this perturbation-theory calculation beyond leading order naturally results in corrections at the single- and two-body level. In addition, it yields genuine three- and multi-atom interactions at higher perturbative orders. 
For instance, we can consider the three-body interaction due to the shift of the three-atom state 
$\rvert sss\rangle \rightarrow \rvert s p p \rangle \rightarrow \rvert p s p \rangle \rightarrow \rvert s s s \rangle $, 
which arises at third order. We present a detailed analysis of the many-body perturbative expansion for a simplified model in the Supplementary Material (SM)~\cite{supp}. The contributions at each successive order are generically suppressed by a factor of the coupling ratio $\sim {W}/{\delta}$, where $W$ is the relevant matrix element of the dipole-dipole interaction operator, leading to fast convergence of the series. This is partially counteracted by the rapidly scaling multiplicities, ${N \choose k}$, of these $k$-body terms in an $N$-atom system, which can result in significant contributions for comparatively high densities and lattice connectivities. Nevertheless, in the conventional limit of large interatomic separations, such higher-order interactions play a negligible role.

\textit{Engineering three-body interactions.---}With this background, we now demonstrate three-atom interactions can actually dominate over the binary forces. Figure~\ref{fig:fig2}(a) shows the calculated level diagram of three atoms arranged in the configuration of Fig.~\ref{fig:fig1}, with transitions induced by the dipole-dipole interaction. The van der Waals forces $V^{(2)}$ may be viewed as originating from binary ``hopping'' processes from the target state to one of the states and back, completely leaving one atom out; e.g., $\rvert sss\rangle$\,$\rightarrow \rvert pps\rangle$\,$\rightarrow \rvert sss\rangle$. Third-order interactions stem from transitions between the excited states in the virtual manifold before the system ``returns'' to the target state. 
While such processes are typically suppressed, we now consider blocks of three atoms (Fig.~\ref{fig:fig1}) where an enhanced detuning on one site can yield dominant three-body interactions. In an extended chain, we label the three sublattices as $1,2,3$ and, without loss of generality, assign the earmarked sites the index 3. After detuning the laser on sublattice 3 into resonance with $|\tilde{s}\rangle$, the target state $|ss\tilde{s}\rangle$ can undergo dipole transitions, as illustrated in Fig.~\ref{fig:fig2}. By tuning the energy of $|\tilde{s}\rangle$ on site 3, it can be brought near resonance with a coupled manifold of states, enhancing interactions mediated by $V_{13}$ and $V_{23}$ but not $V_{12}$. Simultaneously, if the dipole interaction $V_{12}$ exceeds $V_{23}$ and $V_{31}$—ensured by a stretched lattice geometry with $D \ll R$—the effective three-atom interaction is further enhanced (green arrows). We note that the degeneracy with $|\tilde{s}ss\rangle$ and $|s\tilde{s}s\rangle$ is lifted, so the schematic in Fig.~\ref{fig:fig2}(b) provides a faithful approximation to the actual level structure.

This process can be understood in a dressed-state picture. The unperturbed states $|psp\rangle$ and $|spp\rangle$ are degenerate at large particle separations. Since they are directly coupled via the dipole-dipole interaction, they quickly split into a symmetric and an antisymmetric pair state. When either potential energy surface crosses the target state $|ss\tilde{s}\rangle$, the dipole-dipole interactions $V_{13}$ and $V_{23}$ couple the states, leading to an avoided crossing. Note that this avoided crossing is a pure three-body effect and would be entirely absent for binary van der Waals interactions. The splitting is on the order of $V_{13}$ and $V_{23}$.

We now present a numerical example demonstrating the emergence of effective three-body interactions in an array of rubidium atoms. In Fig.~\ref{fig:fig2}(a), the numerically diagonalized three-body potential energy surfaces show how the initially degenerate excited states indeed split and form an anticrossing with the target state.   To keep the energy scales involved reasonable, we propose to excite to the target state $|37S_{-1/2}, 37S_{-1/2},38S_{-1/2} \rangle$ that offers a near-zero F\"orster defect in the transition $37S \leftrightarrow 38P$~\cite{ryabtsev2010}. 
This can be achieved by initializing the atoms in a hyperfine state $|g\rangle$ with dipole coupling to the Rydberg states, and by modulating the involved excitation laser on site(s) 3. For the discussion of the many-body phases below, we collectively refer to the Rydberg states as $|r\rangle$ ($= \rvert \tilde{s}\rangle$ for sites on sublattice 3, and $= \rvert s\rangle$ otherwise). We further consider an optical AC-Stark shift  of $62$ MHz, that enables us to tune the energy of the target state relative to the coupled F\"orster states, and a magnetic field of  $m_J \cdot 20$ MHz  to split the Zeeman levels. 
It is clear that the 
full solution (gray and green lines) strongly deviates from the sum of binary interactions (blue dashed lines) around the point of the avoided crossing.

\begin{figure}[t]
\begin{center}
 \includegraphics[width=0.95\linewidth]{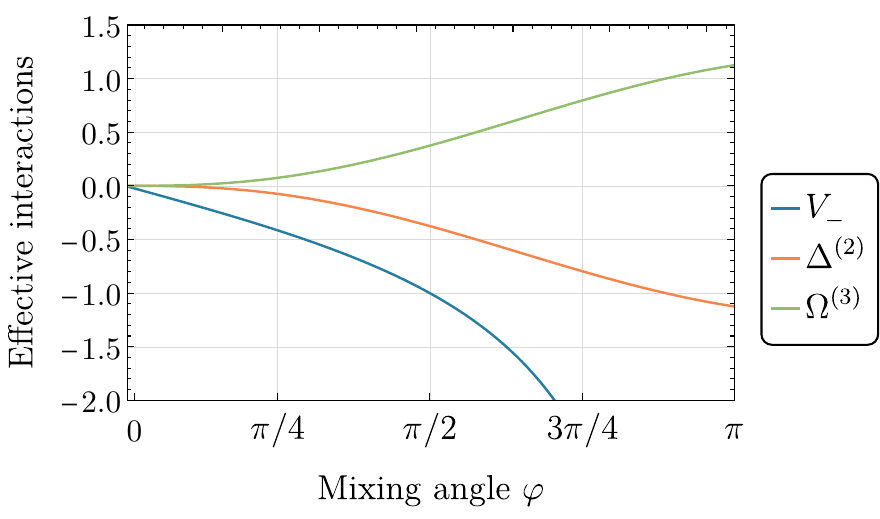}
\end{center}
\caption{Evolution of the coefficients of the three-body terms in the effective Hamiltonian~\eqref{eq:plaquettedef} with the mixing angle $\varphi$. Here, for simplicity, we set $\Delta_{12} = \Delta_{23} = \Delta_{31} \equiv \Delta^{(2)}$, and  $\Omega_{123} = \Omega_{231} = \Omega_{312} \equiv \Omega^{(3)}$. All effective interactions vanish at $\varphi = 0$ and monotonically increase with the mixing angle as the avoided crossing is traversed. At $\varphi = \pi/2$, the triply excited subspace is diagonalized by the equal-superposition states $\rvert \pm \rangle =  (\rvert {\mathcal{R}}\rangle \pm \rvert \bar{\mathcal{R}}\rangle)/\sqrt{2}$. Note that $V_{-} <0$ and $\Delta^{(2)}<0$ contribute to attractive and repulsive three-particle interactions, respectively. }
\label{fig:fig3}
\end{figure}

\textit{Many-body Hamiltonian.---}To illustrate the physical effects of this three-body interaction, we now map it onto a qubit basis. In a direct implementation, such a basis would feature the atomic ground state $|g\rangle$ at each site and its corresponding sublattice-dependent Rydberg level, $|r\rangle$. However, one must keep in mind that when the potential surfaces of the target state ($\rvert \mathcal{R}\rangle \equiv \rvert ss\tilde{s}\rangle$)  and the other triply excited Rydberg state ($\rvert \bar{\mathcal{R}}\rangle \equiv \rvert psp\rangle\pm\rvert spp\rangle$) cross, both are, in general, important. In addition, there is a strong mixing of the electronic character of the adiabatic states, which is parametrized by a mixing angle $\varphi$ (see Eq.~(S27)~\cite{supp}). Hence, we carry out the mapping to an effective qubit basis by elimination of the Rydberg state higher in energy, assuming that it is sufficiently far separated. As detailed in the SM~\cite{supp}, this elimination yields an effective qubit description with a three-body interaction, but, in addition,  the effective Hamiltonian carries not only the standard excitation terms, $\Omega \sigma^x_i$, but also nonlinear excitation terms of the form $\Omega \sigma^{x}_1 n^{}_2n^{}_3$, $\Omega \sigma^x_2 n^{}_1n^{}_3$, etc., involving the Rydberg state projectors $n_i \equiv \rvert r_i \rangle \langle r_i \lvert$ on site $i$.

\begin{figure*}[t]
\begin{center}
 \includegraphics[width=\linewidth]{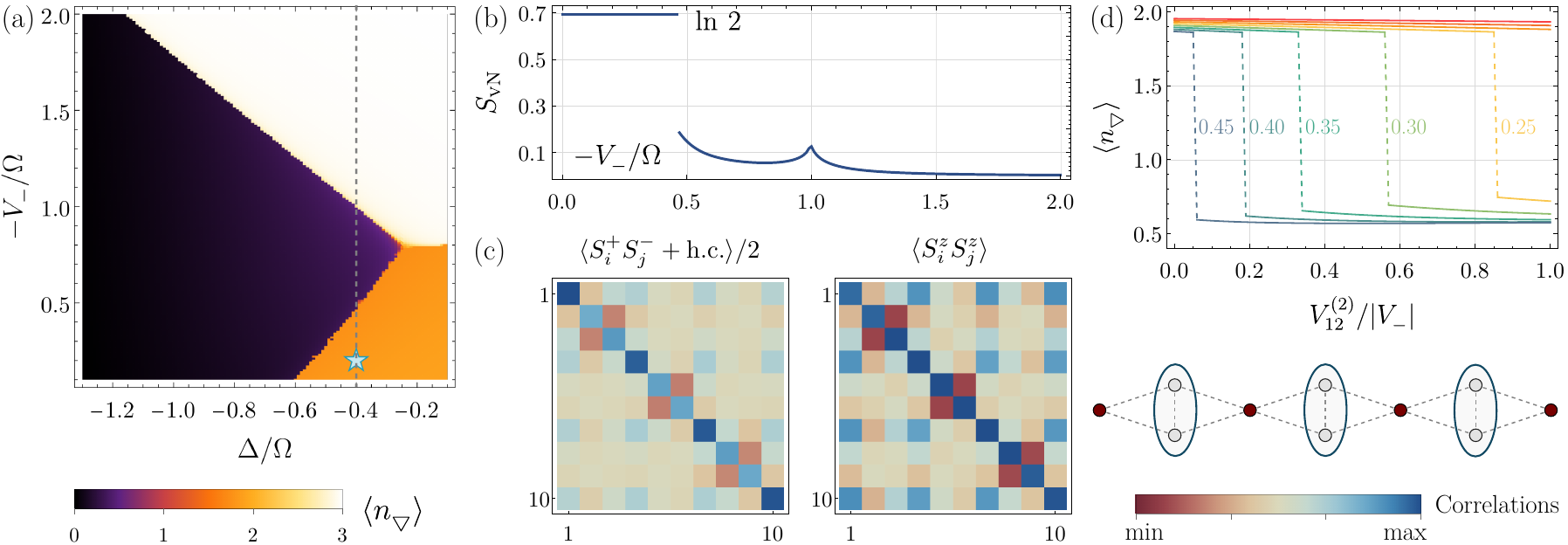}
\end{center}
\caption{(a) Quantum phase diagram with $V^{(2)}_{ij} = 0$ at a mixing angle of $\varphi = \pi/2$, as calculated using DMRG on a quasi-1D chain of 100 sites. (b) The von Neumann entanglement entropy, $S_{\mathrm{vN}}$, as a function of $V_{-}$ along the vertical dashed line in (a). The saturation value of $\ln 2$ indicates that the bipartition of the system cuts across a spin singlet. (c) Transverse ($\langle S_i^+S_j^- + \mathrm{h.c.} \rangle/2$) and longitudinal ($\langle S^z_i S^z_j \rangle$) spin-spin correlation functions calculated at the point in the rung-singlet phase, $\Delta/\Omega=-0.4$, $V_{-}=-0.2$, marked by the star in (a). Here, we display the correlations for the ten sites (corresponding to three full plaquettes) at the center of the 100-site chain. These correlations reveal the structure of the ground state as pictorially depicted, where sites with $n_i =1 $ are marked in red while the blue ellipses denote singlets. 
(d) Focusing on a fixed value of $\Delta/\Omega=-0.4$, we vary the relative strength of the two-body interactions  for different $-V^{}_{-}/\Omega = 0.10$ (red), $0.15, 0.20, \ldots, 0.45$ (blue)  inside the rung-singlet phase, which survives over an extended range of $V^{(2)}_{12}/\lvert V^{}_{-}\rvert$.}
\label{fig:fig4}
\end{figure*}
 
Putting together all these pieces, the full many-body Hamiltonian can therefore be written in the form $H^{}_\text{eff} = H^{(0)}_\text{eff} + \sum_{\bigtriangledown} H_\text{eff}^\bigtriangledown$,  where
\begin{equation}
\label{eq:H0}
    H^{(0)}_\text{eff} = \frac{\Omega}{2} \sum_i\sigma^x_{i} - \Delta \sum_i n^{}_{i} + \sum_{i \leq j} V^{(2)}_{ij} n^{}_{i}n^{}_{j},
\end{equation}
and $\Delta$ and $\Omega$ are the laser detuning and the Rabi frequency, respectively. The three-particle terms acting on each triangular plaquette (denoted by $\bigtriangledown$) are:
\begingroup
\allowdisplaybreaks
\begin{align}
\label{eq:plaquettedef}
& H_\text{eff}^\bigtriangledown =  V^{}_- n^{}_{1}n^{}_{2}n^{}_{3}\\
\nonumber&  + \Delta^{}_{12} n^{}_{1} n^{}_{2}(1-n^{}_{3}) + \Delta^{}_{23} n^{}_{2} n^{}_{3}(1-n^{}_{1})
 + \Delta^{}_{31} n^{}_{3} n^{}_{1}(1-n^{}_{2})\\
 \nonumber
\nonumber &+ \frac{\Omega}{2} \left( \cos \left( \varphi/2 \right) -1 \right)  \left[ \sigma^x_{1}n^{}_{2}n^{}_{3} + \sigma^x_{2}n^{}_{3}n^{}_{1} + \sigma^x_{3}n^{}_{1}n^{}_{2} \right] \\
 \nonumber&+
 \left[\Omega^{}_{123}\sigma^+_{3}\sigma^-_{1}n^{}_{2} 
 +\Omega^{}_{231}\sigma^+_{1}\sigma^-_{2}n^{}_{3} 
 + \Omega^{}_{312}\sigma^+_{2}\sigma^-_{3}n^{}_{1}+ \text{h.c.}\right].
\end{align}
\endgroup
The strengths of the various parameters appearing in this Hamiltonian are plotted in Fig.~\ref{fig:fig3} as a function of $\varphi$.
In the rest of the manuscript, we will focus on the effective Hamiltonian defined by the sum of Eqs.~\eqref{eq:H0} and \eqref{eq:plaquettedef}.

\textit{Effects of three-body terms.---}To explore the new physics induced by the three-body interactions, we systematically examine the quantum phases of the effective model defined above. Specifically, we study the phase diagram of $H_\text{eff}$ numerically, using the density-network renormalization group (DMRG) algorithm to compute matrix product state representations of the ground state. Throughout, we consider an arrangement of atoms in the quasi-1D geometry sketched in Fig.~\ref{fig:fig1}, in which a single atom alternates with a doublet of atoms~\cite{buca2025quantum}. 
A key feature of this geometry (as mentioned above) is that it permits a regime---in the vicinity of the avoided crossing---wherein two-body interactions are much weaker than the three-body ones, enabling us to isolate and identify the effects of the latter.

Before proceeding to the three-body physics, it is useful to recall the expectations for the phase diagram if one had only two-body and onsite terms~\cite{Schiffer2024,lukin2024quantum}. For large, negative $\Delta/\Omega$  and a finite positive two-body interaction $V^{(2)}$, it is energetically favorable for all of the atoms to occupy their atomic ground state $\vert g \rangle$, so the average density of Rydberg excitations per triangular plaquette is $\langle n_{\bigtriangledown} \rangle \approx 0$. On the other hand, for  positive $\Delta/\Omega \lesssim V^{(2)} $, the system transitions to a state where there is one excitation per plaquette, $\langle n_{\bigtriangledown} \rangle$\,$\approx$\,$1$. If the boundary conditions are such that the chain is terminated with 3-sublattice sites, then the state is uniquely determined and the atoms in each doublet remain in $\vert g \rangle$. Eventually for $\Delta/\Omega \gg V^{(2)} $, the energetic penalties of the two-body interactions are overridden and we end up in a fully occupied state with $\langle n_{\bigtriangledown} \rangle \approx 3$. A similar set of phases  with $\langle n_{\bigtriangledown} \rangle \approx 0, 3$ also arises for  attractive two-body interactions and the phase diagram for $V^{(2)} <0 $ is discussed in Sec.~IIB of the SM.

Against this backdrop, we now highlight the modifications caused by the three-body terms. 
To begin, we first set the two-body interactions to zero; the resultant phase diagram is presented in Fig.~\ref{fig:fig4}(a). Interestingly,
we find the emergence of a new phase which was previously absent. This is a purely quantum phase---without a classical ($\Omega, \Omega_{ijk}$\,$=$\,$0$) counterpart---in which $\langle n_{\bigtriangledown} \rangle \approx 2$. Here, the ground state consists of the non-doublet sites being excited to the Rydberg states while a single Rydberg excitation is delocalized on the two other sites of the triangular plaquette as a spin singlet. This structure is reflected both in the von Neumann entanglement entropy across a doublet bipartition, which equals $\ln 2$ for a singlet state [Fig.~\ref{fig:fig4}(b)], and in the spin-spin correlations [Fig.~\ref{fig:fig4}(c)]. 

A natural next question to then ask is whether this quantum ``rung-singlet'' phase~\cite{schmidt2003rung} is special to the fine-tuned limit of only three-body interactions or whether it remains stable once the van der Waals terms are included as well. We answer this question in Fig.~\ref{fig:fig4}(d), which showcases the full phase diagram after incorporating two-body interactions. Labeling these couplings by the sublattices they connect, we recall that  $V^{(2)}_{12} \gg V^{(2)}_{23}, V^{(2)}_{31}$, and thus,  set the latter two to zero. We observe that the rung-singlet phase survives robustly over an extended range of interaction strengths $V^{(2)}_{12}/\lvert V^{}_{-}\rvert \ne 0$. 

\begin{figure}[t]
\begin{center}
 \includegraphics[width=\linewidth]{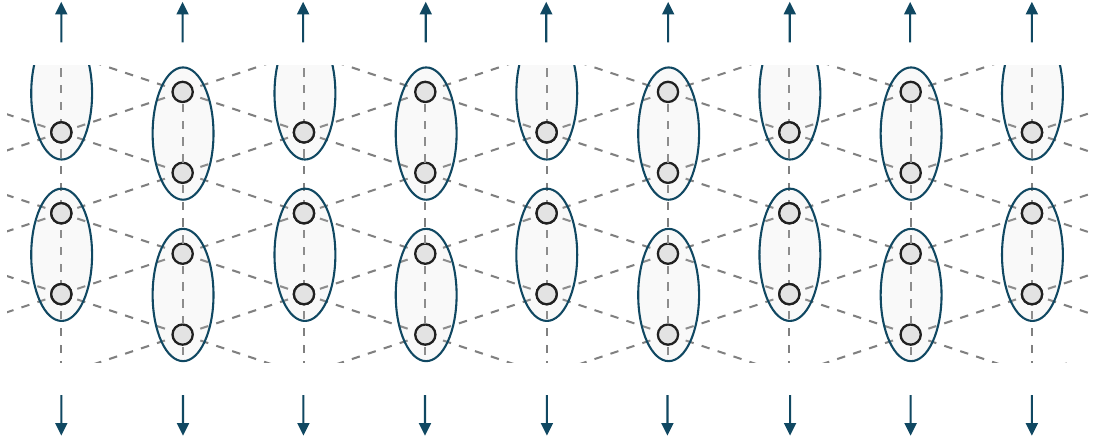}
\end{center}
\caption{Schematic illustration of the 2D analog of the rung-ordered phase stabilized by three-body interactions. On the distorted triangular lattice formed by stacking quasi-1D plaquettes, the singlets (depicted by gray ellipses) can organize into a valence bond solid at $\langle n_{\bigtriangledown} \rangle = 1/4$. The arrows indicate that allowing the dimers along each column to be translated up or down without energetic cost leads to exponentially many such configurations.}
\label{fig:fig5}
\end{figure}

These considerations demonstrate that the inclusion of three-body terms yields physics distinct from that accessible with two-body interactions alone.  Harnessing these novel interactions opens the door to more exotic phenomenology in higher spatial dimensions. For example, one can consider stacking the quasi-1D chains in two dimensions forms a distorted triangular lattice. Suppose that by tuning $\Delta/\Omega$, the Rydberg excitation density is set to $\langle n_{\bigtriangledown} \rangle = 1/4$. Building on the 1D insight that three-body interactions favor singlet formation, the 2D system in this regime can be argued to realize an ordered valence bond solid state, in which the lattice is tiled by a dimer covering of singlets  (Fig.~\ref{fig:fig5})~\cite{moessner2001ising,samajdare2021}. The introduction of additional (suitably chosen) two-body interactions,  that generate resonances between these dimer configurations, can then potentially drive the system into a highly entangled quantum spin liquid~\cite{samajdar2023,semighini2021}. A comprehensive study of these effects in two dimensions constitutes an interesting direction for future work.

\textit{Discussion and outlook.---}In this work, we develop a new component for the Rydberg analog simulation toolbox: the ability to construct and control three-body interactions in lattice geometries. Leveraging experimentally accessible~\cite{Manovitz_2025,hirsbrunner2025} local detunings, our approach expands the range of quantum many-body Hamiltonians simulable with Rydberg atom arrays to beyond the conventional regime dominated by two-body interactions.

A recent approach to Hamiltonian engineering that has shown particular promise for Rydberg systems is Floquet driving~\cite{scholl2022,kunimi2024,tian2025}. In contrast, the method presented here offers a static route for realizing three-body terms, avoiding the heating effects that inevitably accompany periodic driving. Alternatively, digital quantum simulation can be used to engineer three-body interactions~\cite{Lukin-Floquet-1,nishad2023a,maskara2025programmable}, in which the terms on each plaquette are implemented through a programmed gate sequence; however, this approach requires a number of gates that scales linearly with system size. By contrast, our analog framework offers a direct route to generating the full set of desired multibody interactions in a single step, relying solely on the native Hamiltonian.

The controlled implementation of three-body terms enables the study of previously inaccessible models of many-body physics where strongly constrained interactions play a central role~\cite{buechler2007,peng2009,mazza2010,zhang2011,daley2014,hafezi2014,petrov2014}.
This includes, for example, the equilibrium and dynamical properties of effective gauge theories induced by multibody constraints. Going forward, our methodology can also be extended to higher-order interactions, potentially enabling the quantum simulation of models with four-body---or more complex---coupling structures (such as ring exchanges~\cite{moessner2008quantum}) and genuine multipartite entanglement.  The enhanced interqubit connectivity offered by multibody terms in the Hamiltonian would also be of utility for quantum computational applications~\cite{locher2025}. 
For instance,  these novel interactions could potentially be used for new approaches to efficiently generate 
$\mathcal{N}$-Toffoli gates. 
Such capabilities would open up new directions for studying a wide class of 
emergent many-body phenomena and novel approaches to quantum simulations~\cite{hu2026universal}.

\begin{acknowledgments}
\textit{Acknowledgments.---}We thank Dolev Bluvstein, Simon Hollerith, and Thomas Pohl for valuable discussions. 
The calculations presented in this article were implemented using the \texttt{pairinteraction}~\cite{weber2017} software and the \textsc{ITensor} library~\cite{itensor}. The DMRG simulations were performed on  computational resources managed and supported by Princeton Research Computing, a consortium of groups including the Princeton Institute for Computational Science and Engineering (PICSciE) and the Office of Information Technology's High Performance Computing Center and Visualization Laboratory at Princeton University.
This work has been supported by the National Science Foundation under Award $\#$PHYS-2409630 and through a grant for the Institute for Theoretical Atomic, Molecular, and Optical Physics at Harvard University and the Smithsonian Astrophysical Observatory. R.S. was supported by the Princeton Quantum Initiative Fellowship. This work was completed at the Aspen Center for Physics, which is supported by a grant from the Simons Foundation (1161654, Troyer).
\end{acknowledgments}

\bibliographystyle{apsrev4-2_custom}
\bibliography{refs}

\newpage


\section*{Supplementary material}

# Supplemental Material: Three-body interactions in Rydberg lattices

## I. EMERGENT MULTIBODY INTERACTIONS IN RYDBERG SYSTEMS

The total energy of a system with dipolar interactions can be expressed as a perturbative series in the interaction strength: $E^{(1,...,N)} = \sum_i E^{(i)} + \sum_{\langle ij \rangle} \Delta E^{(i,j)} + \sum_{\langle ijk \rangle} \Delta E^{(i,j,k)} + \ldots$, where we decompose the energy into a summation over one-body terms, two-body terms, and so on. Each term in this expansion decays with a characteristic power law with the particle separation: the $p$th-order contribution scales as $R^{-3p}$, where $R$ denotes a typical interparticle distance. Although higher-order terms generally become progressively smaller due to this power-law suppression, their cumulative effect can nonetheless be significant when the number of interacting atoms $N$ is large. Indeed, the number of $p$-body terms, given by $\binom{N}{p}$, increases rapidly with $p$ for $p \ll N$. In this section, we investigate the impact of such multibody interactions within a simplified two-level model of the dipolar Hamiltonian

$$H_d = J \sum_i \sigma_i^+ \sigma_i^- + \frac{|d|^2}{4\pi\varepsilon_0} \sum_{i<j} \frac{1}{R_{ij}^3} (\sigma_i^+ + \sigma_i^-)(\sigma_j^+ + \sigma_j^-), \tag{S1}$$

where $d$ denotes the dipole transition moment. Hereafter, we set the unit of energy to $J$ and the unit of distance to $R_0$, such that the dipolar coupling satisfies $|d|^2/(4\pi\varepsilon_0 R_0^3) = J$. The rescaled Hamiltonian then depends on only a single parameter:

$$H_d = \sum_i \sigma_i^+ \sigma_i^- + \sum_{i<j} \frac{1}{R_{ij}^3} (\sigma_i^+ + \sigma_i^-)(\sigma_j^+ + \sigma_j^-). \tag{S2}$$

The exact binary interaction is given by $V_{ij} = 1 - \sqrt{1 + R_{ij}^{-6}}$, with the perturbative van der Waals limit approximated as $V_{ij} = -1/(2R_{ij}^6)$.

We now consider several lattice geometries of $N$ atoms: a linear chain, a ring, and a square lattice. In each case, we fix the nearest-neighbor distance $a$ such that the relative deviation $\mathcal{E}_b$ between the exact binary interaction and its van der Waals approximation is

$$\mathcal{E}_b = -\frac{(1/a)^6}{2\left(1 - \sqrt{1 + (1/a)^6}\right)} - 1 \quad \Leftrightarrow \quad a = \left[4(\mathcal{E}_b + \mathcal{E}_b^2)\right]^{-\frac{1}{6}}. \tag{S3}$$

For all numerical examples, we fix $\mathcal{E}_b = 1\%$, corresponding to $a = 1.71$. Since all other interatomic distances are larger than $a$, the total error of the binary approximation with respect to the full many-body interaction is bounded by $\mathcal{E}_b$. Any additional deviations thus have to necessarily arise from genuine many-body effects involving three or more particles. In the following, we summarize a systematic perturbative approach based on Van Vleck theory and compare its predictions with exact numerical results (Fig. S2).

The dipole-dipole interaction couples the ground state(s) of the noninteracting system—referred to as the "model space"—to excited states outside this space. Since this coupling is typically weak, we employ Van Vleck perturbation theory to construct a transformation which eliminates it, yielding an effective Hamiltonian that acts only within the model space. To this end, we partition the full Hilbert space into two sets of many-body states: the model space $\alpha$ (which may consist of a single state in the simplest case), and its complement $\beta$. Any operator $A$ can then be decomposed into diagonal and off-diagonal parts with respect to this partition:

$$A = A_D + A_X, \tag{S4}$$
$$A_D = P_\alpha A P_\alpha + P_\beta A P_\beta, \qquad A_X = P_\alpha A P_\beta + P_\beta A P_\alpha, \tag{S5}$$

where $P_\alpha$ and $P_\beta$ are the projectors onto the respective subspaces. In particular, the noninteracting Hamiltonian $H_0$ is purely diagonal ($H_0 = H_D$), while the interaction $V$ generally consists of both diagonal and off-diagonal components, i.e., $V = V_D + V_X$. Following the perturbative expansion outlined in Ref. [1], we derive a general expression for the effective Hamiltonian up to fourth order and then apply this framework to the dipole-dipole interaction. The resulting effective Hamiltonian acting on subspace $\alpha$ has the form

$$H_{\text{eff}} = P_\alpha H_0 P_\alpha + P_\alpha W P_\alpha, \tag{S6}$$

where the operator $W$ encapsulates the interaction effects but no longer couples the $\alpha$ and $\beta$ subspaces.

## A. Van Vleck perturbation theory

The general expansion is based on an iterative solution of the decoupling transformation, which is expressed in terms of an operator $G$ [1] and expanded in orders of the interaction. This transformation is defined to act only between different subspaces, i.e., $G = G_X$. From the power-series expansion, we obtain the following expression up to third order:

$$G^{(0)} = 0, \tag{S7}$$

$$\left[H_0, G^{(1)}\right] = -V_X, \tag{S8}$$

$$\left[H_0, G^{(2)}\right] = -\left[V_D, G^{(1)}\right], \tag{S9}$$

$$\left[H_0, G^{(3)}\right] = -\left[V_D, G^{(2)}\right] - \frac{1}{3}\left[\left[V_X, G^{(1)}\right], G^{(1)}\right]. \tag{S10}$$

Since $G$ acts only *between* different subspaces whereas $H_0$ acts solely *within* each subspace, we can directly find the matrix elements of the above equations and hence, $G$. We introduce $\gamma, \bar{\gamma} \in \{\alpha, \beta\}$, $\gamma \neq \bar{\gamma}$, such that

$$\langle i, \gamma|[H_0, G^{(1)}]|j, \bar{\gamma}\rangle = -\langle i, \gamma|V_X|j, \bar{\gamma}\rangle \implies \langle i, \gamma|G^{(1)}|j, \bar{\gamma}\rangle = \frac{\langle i, \gamma|V_X|j, \bar{\gamma}\rangle}{E_{j,\bar{\gamma}} - E_{i,\gamma}}. \tag{S11}$$

Using the shorthand notation $\langle i, \gamma|O|j, \bar{\gamma}\rangle \equiv O^{i,\gamma}_{j,\bar{\gamma}}$, a similar approach as above works for the higher-order terms

$$\langle i, \gamma|G^{(2)}|j, \bar{\gamma}\rangle = \frac{\langle i, \gamma|[V_D, G^{(1)}]|j, \bar{\gamma}\rangle}{E_{j,\bar{\gamma}} - E_{i,\gamma}} = \frac{(V_D)^{i,\gamma}_{k,\gamma}(G^{(1)})^{k,\gamma}_{j,\bar{\gamma}} - (G^{(1)})^{i,\gamma}_{k,\bar{\gamma}}(V_D)^{k,\bar{\gamma}}_{j,\bar{\gamma}}}{E_{j,\bar{\gamma}} - E_{i,\gamma}}, \tag{S12}$$

$$\begin{aligned}\langle i, \gamma|G^{(3)}|j, \bar{\gamma}\rangle &= \frac{\langle i, \gamma|[V_D, G^{(2)}]|j, \bar{\gamma}\rangle}{E_{j,\bar{\gamma}} - E_{i,\gamma}} + \frac{\langle i, \gamma|[[V_X, G^{(1)}], G^{(1)}]|j, \bar{\gamma}\rangle}{3(E_{j,\bar{\gamma}} - E_{i,\gamma})}, \\ &= \frac{(V_D)^{i,\gamma}_{k,\gamma}(G^{(2)})^{k,\gamma}_{j,\bar{\gamma}} - (G^{(2)})^{i,\gamma}_{k,\bar{\gamma}}(V_D)^{k,\bar{\gamma}}_{j,\bar{\gamma}}}{E_{j,\bar{\gamma}} - E_{i,\gamma}} \\ &+ \frac{(V_X)^{i,\gamma}_{k,\bar{\gamma}}(G^{(1)})^{k,\bar{\gamma}}_{k',\gamma}(G^{(1)})^{k',\gamma}_{j,\bar{\gamma}} + (G^{(1)})^{i,\gamma}_{k,\bar{\gamma}}(G^{(1)})^{k,\bar{\gamma}}_{k',\gamma}(V_X)^{k',\gamma}_{j,\bar{\gamma}} - 2(G^{(1)})^{i,\gamma}_{k,\bar{\gamma}}(V_X)^{k,\bar{\gamma}}_{k',\gamma}(G^{(1)})^{k',\gamma}_{j,\bar{\gamma}}}{3(E_{j,\bar{\gamma}} - E_{i,\gamma})}.\end{aligned} \tag{S13}$$

Here and in the following, repeated Roman indices are implicitly summed over. All of these expressions are thus determined by $V_X$, $V_D$, and $H_0$, as expected. The decoupled effective Hamiltonian can now be expressed in terms of the transformation $G$, again proceeding order-by-order:

$$W^{(1)} = V_D, \tag{S14}$$

$$W^{(2)} = \frac{1}{2}\left[V_X, G^{(1)}\right], \tag{S15}$$

$$W^{(3)} = \frac{1}{2}\left[V_X, G^{(2)}\right], \tag{S16}$$

$$W^{(4)} = \frac{1}{2}\left[V_X, G^{(3)}\right] - \frac{1}{24}\left[\left[\left[V_X, G^{(1)}\right], G^{(1)}\right], G^{(1)}\right]. \tag{S17}$$

Next, we immediately project into the model space $\alpha$, which is our main object of focus:

$$\left(W^{(1)}\right)^{i,\alpha}_{j,\alpha} = (V_D)^{i,\alpha}_{j,\alpha}, \tag{S18}$$

$$\left(W^{(2)}\right)^{i,\alpha}_{j,\alpha} = \frac{1}{2}\left((V_X)^{i,\alpha}_{k,\beta}(G^{(1)})^{k,\beta}_{j,\alpha} - (G^{(1)})^{i,\alpha}_{k,\beta}(V_X)^{k,\beta}_{j,\alpha}\right), \tag{S19}$$

$$\left(W^{(3)}\right)^{i,\alpha}_{j,\alpha} = \frac{1}{2}\left((V_X)^{i,\alpha}_{k,\beta}(G^{(2)})^{k,\beta}_{j,\alpha} - (G^{(2)})^{i,\alpha}_{k,\beta}(V_X)^{k,\beta}_{j,\alpha}\right), \tag{S20}$$

$$\begin{aligned}\left(W^{(4)}\right)^{i,\alpha}_{j,\alpha} &= \frac{1}{2}\left((V_X)^{i,\alpha}_{k,\beta}(G^{(3)})^{k,\beta}_{j,\alpha} - (G^{(3)})^{i,\alpha}_{k,\beta}(V_X)^{k,\beta}_{j,\alpha}\right) - \frac{1}{24}(V_X)^{i,\alpha}_{k,\beta}(G^{(1)})^{k,\beta}_{k',\alpha}(G^{(1)})^{k',\alpha}_{k'',\beta}(G^{(1)})^{k'',\beta}_{j,\alpha} \\ &+ \frac{3}{24}(G^{(1)})^{i,\alpha}_{k,\beta}(V_X)^{k,\beta}_{k',\alpha}(G^{(1)})^{k',\alpha}_{k'',\beta}(G^{(1)})^{k'',\beta}_{j,\alpha} - \frac{3}{24}(G^{(1)})^{i,\alpha}_{k,\beta}(G^{(1)})^{k,\beta}_{k',\alpha}(V_X)^{k',\alpha}_{k'',\beta}(G^{(1)})^{k'',\beta}_{j,\alpha} \\ &+ \frac{1}{24}(G^{(1)})^{i,\alpha}_{k,\beta}(G^{(1)})^{k,\beta}_{k',\alpha}(G^{(1)})^{k',\alpha}_{k'',\beta}(V_X)^{k'',\beta}_{j,\alpha}.\end{aligned}$$

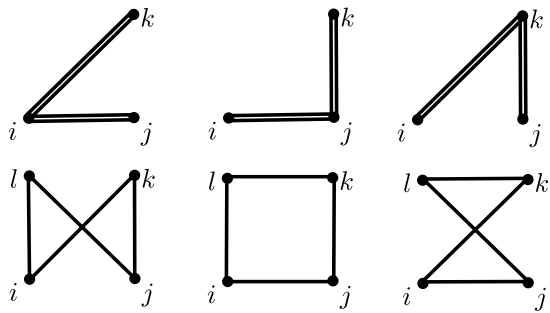

FIG. S1. Fourth-order diagrams contributing to three-particle interactions (top row) and to four-particle interactions (bottom row).

This is the general Van Vleck perturbation series up to fourth order.

### B. Analytic fourth-order effective model for dipolar interactions

We now apply the perturbation theory to the simplified model of interacting two-level atoms, described by Eq. (S2). Carrying out the perturbative expansion to fourth order, as described above, brings us to the effective Hamiltonian

$$\begin{aligned}
H_{\text{eff}}^{(4)} =& \sum_{i<j} \left( -\frac{1}{2R_{ij}^6} + \frac{1}{8R_{ij}^{12}} \right) \mathbb{P}_i \mathbb{P}_j \\
&+ \sum_{i<j<k} \left( \frac{3}{2R_{ij}^3 R_{jk}^3 R_{ik}^3} - \frac{2}{8} \left[ \frac{1}{R_{ij}^6 R_{ik}^6} + \frac{1}{R_{ij}^6 R_{jk}^6} + \frac{1}{R_{ik}^6 R_{jk}^6} \right] \right) \mathbb{P}_i \mathbb{P}_j \mathbb{P}_k \\
&+ \sum_{i<j<k<l} \left( -\frac{20}{8} \right) \left( \frac{1}{R_{ik}^3 R_{jk}^3 R_{jl}^3 R_{il}^3} + \frac{1}{R_{ij}^3 R_{jk}^3 R_{kl}^3 R_{il}^3} + \frac{1}{R_{ij}^3 R_{jl}^3 R_{lk}^3 R_{ik}^3} \right) \mathbb{P}_i \mathbb{P}_j \mathbb{P}_k \mathbb{P}_l,
\end{aligned} \quad \text{(S21)}$$

where $\mathbb{P}_i$ projects onto the lower-energy state of atom $i$. In particular, the fourth-order contributions from three-particle and four-particle interactions are illustrated in Fig. S1. Note that not all possible four-particle diagrams contribute; for instance, disconnected diagrams are absent [2]. In general, these diagrams can be categorized as either irreducible—fully connected—or reducible into lower-order subdiagrams, and this distinction proves useful for constructing approximate higher-order terms.

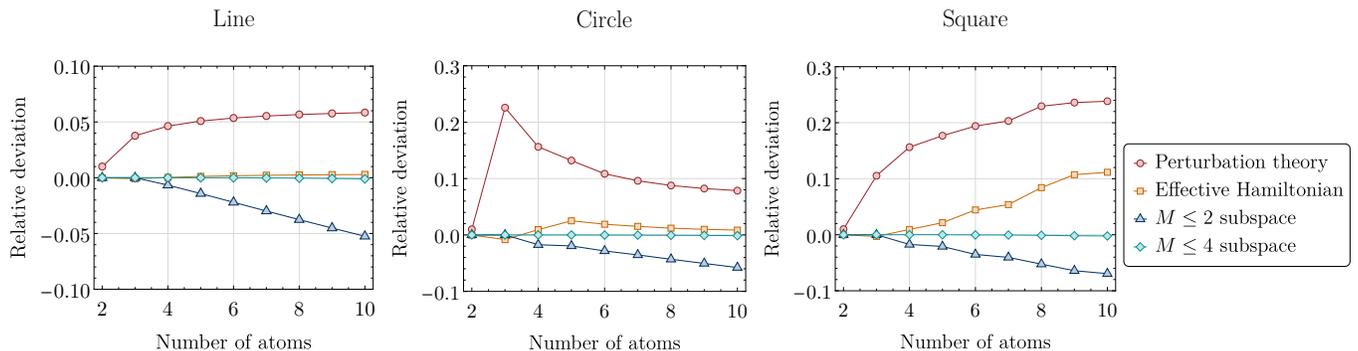

FIG. S2. Relative deviations of the second-order perturbative approximation (additive van der Waals interactions; red), and the effective fourth-order Hamiltonian (orange) from the exact numerical solution as a function of the number of atoms. For comparison, we show the results obtained from diagonalization in a subspace limited to $M \leq 2$ (blue) and $M \leq 4$ (green) excitations. The atoms are arranged regularly on (a) a line, (b) a circle, and (c) a square array; when the number of atoms is not a square number, the "square" geometry is incomplete. The minimal distance between nearest neighbors is chosen such that the nearest-neighbor relative deviation of the perturbative solution is $\mathcal{E}_b = 1\%$.

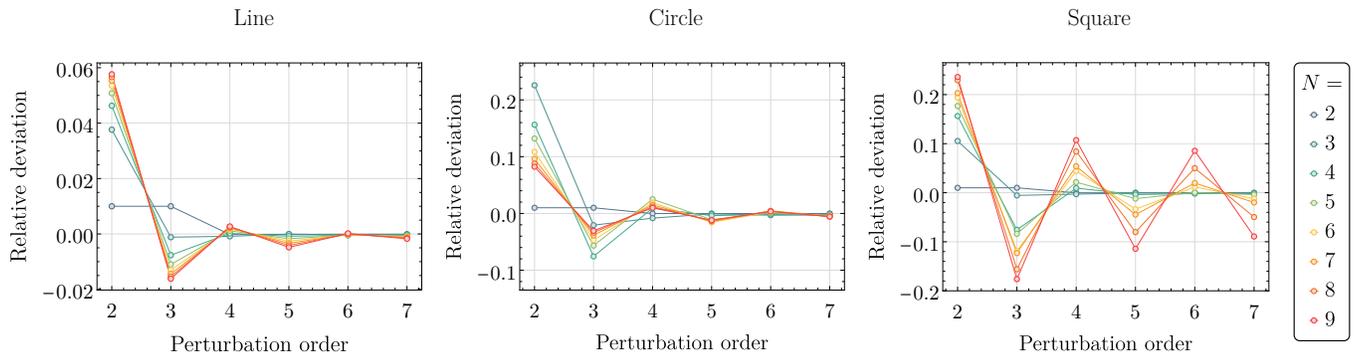

FIG. S3. Convergence of the energy shift, as calculated using an effective many-body Hamiltonian, as a function of the order in perturbation theory. The results converge fast for the line (left) and circle (center) geometries, but require many terms for the square configuration (right).

Finally, we use this effective Hamiltonian to compute the energy shift and provide comparisons to other approximation schemes: our results are shown in Fig. S2. For the linear chain and the regular $N$-gon, we find excellent agreement with the exact solution. In contrast, the approximation shows significant deviations for the "square" configuration, particularly at large $N$.

### C. General-order solution for the many-body dipole interaction

To investigate how convergence in the three configurations is eventually reached, if at all, we numerically continue the Van Vleck perturbation series. Employing the iterative relations arising from the series expansion derived by Shavitt and Redmon [1], we have

$$[H_0, G] = -[V_D, G] - \sum_{p=0}^{\infty} c_p G_p{}^{2p} V_X, \qquad W_C = V_D + \sum_{p=0}^{\infty} t_p \left(\frac{1}{2} G_p\right)^{2p+1} V_X, \tag{S22}$$

and together with the definition $GX = [X, G]$, we can conveniently solve the perturbation series to higher order. The constants $c_p$ and $t_p$ are closely related to the Bernoulli numbers and are defined, e.g., in Ref. 1. The relative deviations thus obtained are plotted in Fig. S3. While we observe rapid convergence for both the linear configuration and the regular $N$-gon, the relative deviation for the square configuration exhibits persistent oscillations around the exact value. We accordingly conclude that many-body corrections become more significant in geometries with a larger number of closely spaced atoms, i.e., a higher coordination number.

## II. THREE-BODY INTERACTIONS IN A QUBIT SYSTEM

In this section, we illustrate the derivation of the effective Hamiltonian in Eqs. (1) and (2) of the main text. This requires mapping the many-body dipolar interactions onto a qubit basis, with two degrees of freedom per site, as we do next.

### A. Effective Hamiltonian

For each plaquette, there are nine relevant states: the ground state $|\mathcal{G}\rangle$ ($|ggg\rangle$), three states with one Rydberg excitation, $|r_i\rangle$, three states with two Rydberg excitations, $|r_{ij}\rangle$, the triply excited state $|\mathcal{R}\rangle$, and another state with three Rydberg excitations, $|\bar{\mathcal{R}}\rangle$. The latter is optically dark but lends the three-particle interactions discussed in this article. The bare-state Hamiltonian (in the rotating frame) reads

$$H_0 = -\Delta \sum_{i=1}^{3} |r_i\rangle\langle r_i| - 2\Delta \sum_{\substack{i,j=1 \\ i\neq j}}^{3} |r_{ij}\rangle\langle r_{ij}| - 3\Delta |\mathcal{R}\rangle\langle \mathcal{R}| - (3\Delta + \bar{\delta})|\bar{\mathcal{R}}\rangle\langle \bar{\mathcal{R}}|, \tag{S23}$$

with $\Delta = \omega_l - \omega_r$ (where $\omega_l$ is the laser frequency and $\omega_r$ is the energy of the relevant Rydberg state), and $\bar{\delta} = \omega_R - \omega_{\bar{R}}$ the relative energy shift between $|\mathcal{R}\rangle$ and $|\bar{\mathcal{R}}\rangle$. The laser driving couples the states as

$$H_l = \frac{\Omega}{2} \sum_{i=1}^{3} |\mathcal{G}\rangle\langle r_i| + \frac{\Omega}{2} \sum_{\substack{i,j=1 \\ i \neq j}}^{3} \left(|r_i\rangle\langle r_{ij}| + |r_{ij}\rangle\langle \mathcal{R}|\right) + \text{h.c.} \quad . \tag{S24}$$

While, in principle, the Rabi frequencies on each site can all be different depending on the local choice of the principal quantum number and the laser intensity, here, we set them to be equal for simplicity. The interaction Hamiltonian is well described by the combination of two-atom interactions, the interaction shift of the state $|\bar{\mathcal{R}}\rangle$, $V_{\bar{\mathcal{R}}}$, and the effective coupling $V > 0$ acting between the triply excited Rydberg states:

$$H_{\text{int}} = \sum_{\substack{i,j=3 \\ i<j}}^{3} V_{ij} |r_{ij}\rangle\langle r_{ij}| + V_{\mathcal{R}} |\mathcal{R}\rangle\langle \mathcal{R}| + V_{\bar{\mathcal{R}}} |\bar{\mathcal{R}}\rangle\langle \bar{\mathcal{R}}| + V(|\mathcal{R}\rangle\langle \bar{\mathcal{R}}| + |\bar{\mathcal{R}}\rangle\langle \mathcal{R}|). \tag{S25}$$

The shift of the triply excited Rydberg state $|\mathcal{R}\rangle$ is well captured by the sum of two-particle interactions as $V_{\mathcal{R}} = V_{12} + V_{23} + V_{13}$.

We now discuss how to reduce this nine-level system to an effective system of three qubits in excellent approximation of the ground-state properties. To do so, we consider the avoided crossing in the two-state subspace of $|\mathcal{R}\rangle$ and $|\bar{\mathcal{R}}\rangle$, where the detuning between $|\mathcal{R}\rangle$ and $|\bar{\mathcal{R}}\rangle$ is $\delta = V_{\bar{\mathcal{R}}} - V_{\mathcal{R}} - \bar{\delta}$ (Fig. S4), and the mixing is fully determined by the angle $\varphi \in (0, \pi)$

$$\tan(\varphi/2) = \frac{-\delta + \sqrt{\delta^2 + 4V^2}}{2V} \Leftrightarrow \tan(\varphi) = \frac{2V}{\delta}. \tag{S26}$$

Consequently, the two hybridized potentials are

$$V_+ = \frac{V}{\tan(\varphi/2)}, \qquad V_- = -V \tan(\varphi/2). \tag{S27}$$

The triply excited subspace is diagonalized via $|-\rangle = \cos(\varphi/2) |\mathcal{R}\rangle - \sin(\varphi/2) |\bar{\mathcal{R}}\rangle$, $|+\rangle = \sin(\varphi/2) |\mathcal{R}\rangle + \cos(\varphi/2) |\bar{\mathcal{R}}\rangle$:

$$\begin{pmatrix} 0 & V \\ V & \delta \end{pmatrix} = \begin{pmatrix} \cos(\varphi/2) & \sin(\varphi/2) \\ -\sin(\varphi/2) & \cos(\varphi/2) \end{pmatrix} \begin{pmatrix} V_- & 0 \\ 0 & V_+ \end{pmatrix} \begin{pmatrix} \cos(\varphi/2) & -\sin(\varphi/2) \\ \sin(\varphi/2) & \cos(\varphi/2) \end{pmatrix}. \tag{S28}$$

We can summarize the rotation of the full Hamiltonian as $H \to U^{-1}HU$ with $U = \mathbb{1} \oplus U_r$, where $U_r$ is the first matrix on the right-hand side of Eq. (S28). Since we are interested in the ground states, we formulate an effective Hamiltonian that considers only the lowest eight states, while treating the coupling to the highest state, $|+\rangle$, perturbatively. The matrix elements of the effective Hamiltonian to second order in the optical coupling are given by [3]

$$\langle i|H_{\text{eff}}|j\rangle = \langle i|H|j\rangle + \frac{\langle i|U^{-1}HU|+\rangle\langle +|U^{-1}HU|j\rangle}{2} \left(\frac{1}{E_i - V_+} + \frac{1}{E_j - V_+}\right). \tag{S29}$$

The Hamiltonian thus obtained is the starting point for the many-body phases investigated in the main text. In qubit notation, it can be decomposed as $H_{\text{eff}} = H_{\text{eff}}^{(0)} + \sum_{\triangledown} H_{\text{eff}}^{\triangledown}$. The standard part

$$H_{\text{eff}}^{(0)} = \frac{\Omega}{2} \sum_i \sigma_i^x - \Delta \sum_i n_i + \sum_{i \leq j} V_{ij} n_i n_j, \tag{S30}$$

describes the familiar laser-excitation, detuning, and two-atom interaction terms. The new three-particle interactions act on each plaquette (denoted by the superscript $\triangledown$):

$$H_{\text{eff}}^{\triangledown} = V_- n_1 n_2 n_3 + \Delta_{12} n_1 n_2 (1 - n_3) + \Delta_{23} n_2 n_3 (1 - n_1) + \Delta_{31} n_3 n_1 (1 - n_2) \tag{S31}$$
$$+ \frac{\Omega}{2} \left(\cos(\varphi/2) - 1\right) \left(\sigma_1^x n_2 n_3 + \sigma_2^x n_3 n_1 + \sigma_3^x n_1 n_2\right) + \left[\Omega_{123} \sigma_3^+ \sigma_1^- n_2 + \Omega_{231} \sigma_1^+ \sigma_2^- n_3 + \Omega_{312} \sigma_2^+ \sigma_3^- n_1 + \text{h.c.}\right]

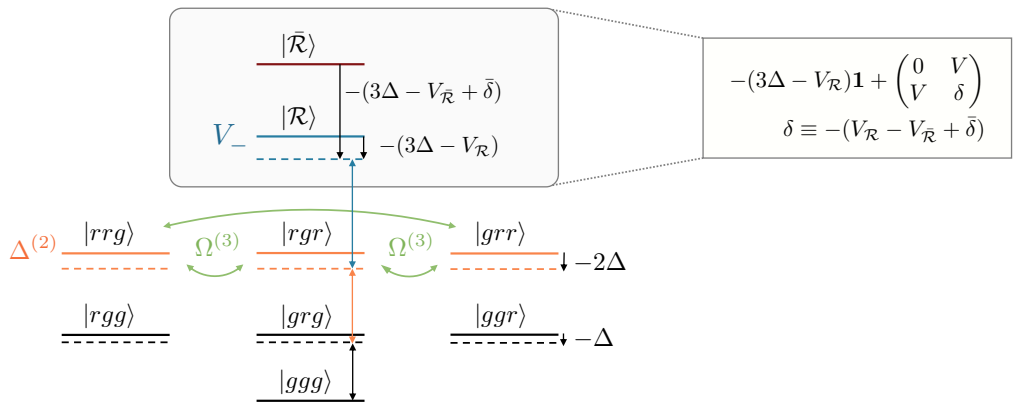

FIG. S4. Elimination of the high-energy state in the avoided crossing yields effective interactions between the atoms. Apart from the native binary (neglected here for simplicity) and three-atom shifts ($V_-$, blue), we obtain an effective energy shift $\Delta^{(2)}$ in the two-excitation subspace, as well as a conditional excitation transfer in the same subspace, denoted by $\Omega^{(3)}$. All levels are detuned by $-\Delta$ for each atom in the Rydberg state and are coupled optically by $\Omega$. The yellow box on the right summarizes the Hamiltonian in the triply excited two-level subspace spanned by $|\mathcal{R}\rangle$ and $|\bar{\mathcal{R}}\rangle$.

with the light-induced terms

$$\begin{aligned}
\Delta_{ij} &= -\frac{\Omega^2(-1+\cos(\varphi))}{8(V_{ij}-V_{\mathcal{R}}+\Delta-V\cot(\varphi/2))} \\
\Omega_{ijk} &= \frac{\Omega^2 \sin(\varphi/2)\left(\sin(\varphi/2)(2\Delta+V_{ij}+V_{jk}-2V_{\mathcal{R}})-2V\cos(\varphi/2)\right)}{8\left(\Delta-V\cot(\varphi/2)+V_{ij}-V_{\mathcal{R}}\right)\left(-\Delta+V\cot(\varphi/2)-V_{jk}+V_{\mathcal{R}}\right)}.
\end{aligned} \qquad (S32)$$

In Eq. (S31), the first term describes the pure three-body interaction. Additionally, note that the two-body terms also acquire an AC-Stark shift from the eliminated state of highest energy, as reflected in the first line. The second line describes laser excitation transfer between the two-excitation and the three-excitation subspaces and hopping conditioned on the presence of another excitation.

### B. Phase diagrams

Starting from the Hamiltonian defined by the sum of Eqs. (S30) and (S31), we construct the phase diagram of the system, shown in Fig. 4(a) of the main text and reproduced here in Fig. S5(a). Throughout this section we focus on the case of mixing angle $\varphi = \pi/2$, and we obtain the corresponding phase boundaries using DMRG simulations on a quasi-one-dimensional chain of 100 sites.

As discussed earlier, the phase diagram showcases three distinct phases. For large negative detuning $\Delta/\Omega$, the many-body ground state consists predominantly of atoms in their electronic ground state $|g\rangle$, yielding an average Rydberg excitation density per triangular plaquette of $\langle n_\triangledown \rangle \approx 0$. This configuration becomes unstable when the attractive three-body interaction $V_-/\Omega$ is sufficiently large and negative, favoring instead fully occupied plaquettes with $\langle n_\triangledown \rangle \approx 3$. The competition between detuning and the three-body attraction shifts the phase boundary between these two regimes leftward with increasing $-V_-$.

Between these two limits of $\langle n_\triangledown \rangle$ lies the rung-singlet phase, which emerges at intermediate values of $\Delta/\Omega$ and $-V_-/\Omega$ and is characterized by an excitation density $\langle n_\triangledown \rangle \approx 2$. In this phase, the non-doublet site of each plaquette is fully excited, while a single Rydberg excitation on the remaining two sites forms a spin singlet that is delocalized across the rung. This phase can be more clearly identified by considering certain additional observables. A particularly convenient quantity in this regard is the average transverse magnetization on a rung, $\langle S_{12}^x \rangle \equiv \langle S_1^x + S_2^x/2 \rangle$. In the paramagnetic regime with $\langle n_\triangledown \rangle \approx 0$, for small $\Delta/\Omega$, the spins are all nearly fully polarized along the $-\hat{x}$ direction, resulting in $\langle S_{12}^x \rangle \to -0.5$. Upon moving deeper into this phase by making $\Delta/\Omega$ more negative, the spins rotate continuously and collectively toward the $-\hat{z}$ direction, which is reflected in a monotonic increase of $\langle S_{12}^x \rangle$. On the other hand, in the rung-singlet phase, one finds $\langle S_{12}^x \rangle = 0$, as expected for a singlet. Moreover, the transverse magnetization on the non-doublet sites, which are nearly fully polarized along $+\hat{z}$, also vanishes, i.e., $\langle S_3^x \rangle = 0$. Thus, unlike in the

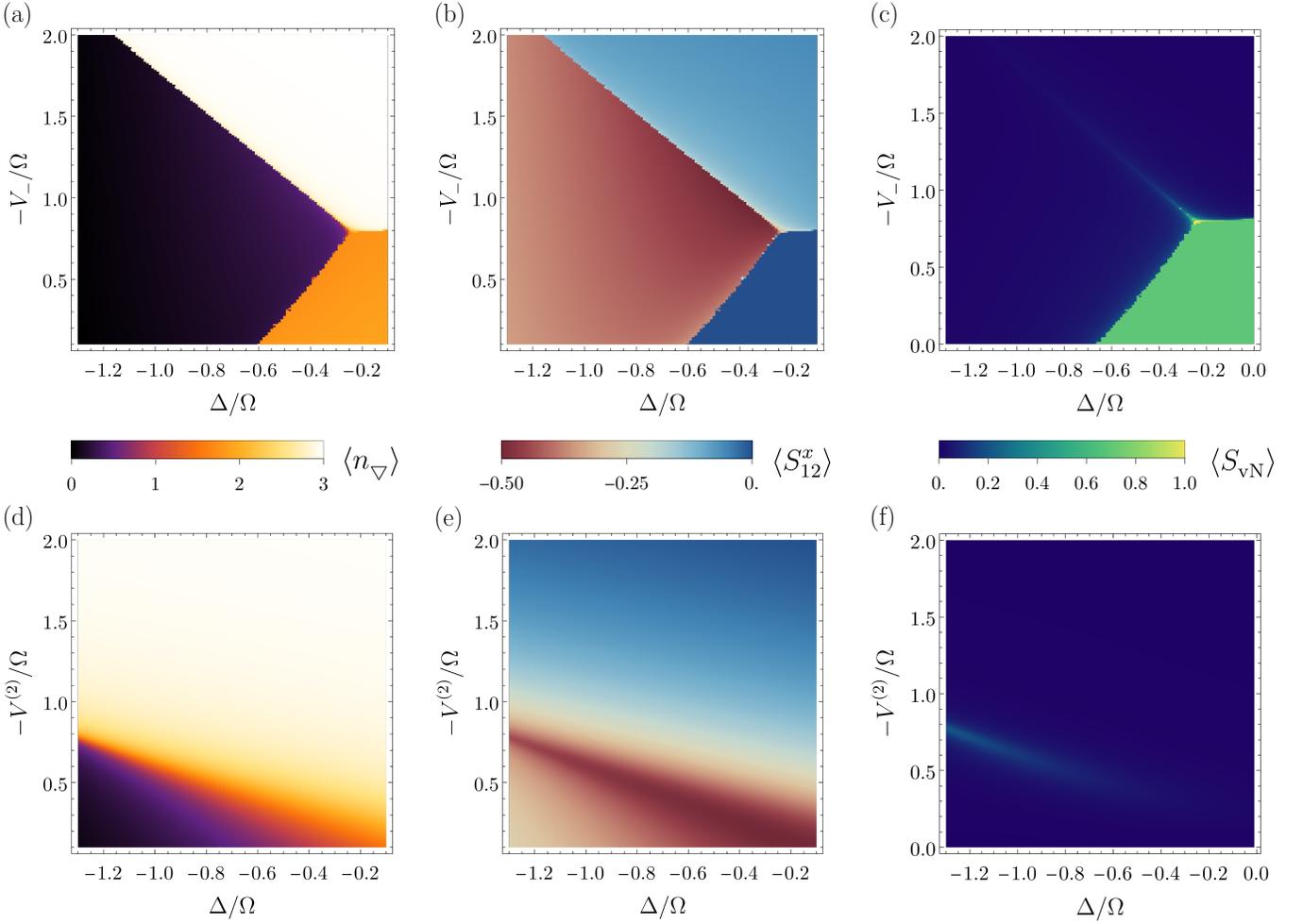

FIG. S5. Phase diagram with three-body interactions based on the (a) plaquette density, (b) the expectation value of $S^x$ averaged over the rungs of the lattice, and (c) the von Neumann entanglement entropy. (d,e,f) Same as above but with only two-body interactions, demonstrating the absence of the rung-singlet phase.

paramagnetic phase—where all three sites of a plaquette are exactly equivalent, share the same orientation on the Bloch sphere, and display identical values of $\langle S^x \rangle$ and $\langle S^z \rangle$—the rung-singlet phase breaks the equivalence between doublet and non-doublet sites.

Beyond these *local* metrics based on spin expectation values, a complementary *global* diagnostic is provided by the von Neumann entanglement entropy $S_{\mathrm{vN}}$, plotted in Fig. S5(c). In the rung-singlet phase, $S_{\mathrm{vN}}$ saturates to $\ln 2$, while it vanishes in the classical phases.

It is instructive to contrast these results with the phase diagram obtained with exclusively two-body interactions, illustrated in Figs. S5(d)–(f). For a direct comparison, we consider the Hamiltonian of Eq. (S30) with an attractive two-body interaction $V^{(2)}$ scanned over the same detuning and interaction ranges as above. In this case, the qualitative behavior is markedly different. As shown in Fig. S5(d), the rung-singlet phase is notably absent: instead of a sharp transition between classical configurations with $\langle n_\triangledown \rangle \approx 0$ and $\langle n_\triangledown \rangle \approx 3$, the system exhibits a smooth crossover, consistent with the fact that neither regime breaks any lattice symmetry, including translations. This interpretation is further supported by the behavior of both $\langle S^x_{12} \rangle$ and $S_{\mathrm{vN}}$, which show no signatures of a distinct intermediate phase (Fig. S5(e-f)).

These observations highlight the crucial role of genuine three-body interactions. The transverse terms, such as $\sigma^x_1 n_2 n_3$, stabilize the structure of the rung-singlet phase in a manner that cannot be replicated by two-body interactions alone. By contrast, purely diagonal three-body terms of the form $n_1 n_2 n_3$ are qualitatively equivalent to combinations of pairwise interactions, $n_1 n_2 + n_2 n_3 + n_3 n_1$, and therefore do not by themselves generate new phases. It is the interplay of transverse and diagonal three-body couplings that fundamentally reshapes the phase diagram and enables the emergence of the rung-singlet state.



\end{document}